\documentclass[a4paper]{article}      
\usepackage{graphicx}
\usepackage{amsmath}
\usepackage{amsfonts}

\begin{document}

\title{\bf{REGULARIZATION AMBIGUITY PROBLEM FOR \\
       THE CHERN-SIMONS SHIFT}%
            \thanks{%
       To be publised in
       \bf{Modern Physics Letter A,
           Vol.13, No.27 (1998) pp.2231-2237.}
        }
       }
\author{Koh-ichi Nittoh%
        \thanks{E-mail: tea@cuphd.nd.chiba-u.ac.jp}\\
\it Graduate School of Science and Technology, Chiba University,\\
\it Chiba 263-8522, Japan \\ \\
       Toru Ebihara%
        \thanks{E-mail: ebihara@cuphd.nd.chiba-u.ac.jp}\\
\it Department of Physics, Faculty of Science, Chiba University,\\
\it Chiba 263-8522, Japan}

\date{July, 1998\\
      CHIBA-EP-86(-REV)}

\maketitle 

\begin{abstract}
We consider the Chern-Simons parameter shift
with the hybrid regularization consisting of the
higher covariant derivative (HCD) and the Pauli-Villars (PV) regulators.
We show that the shift is closely related
to the parity of the regulators 
and get the shift and no-shift results
by a suitable choice of the PV regulators.  
A naive treatment of the HCD term leads incorrect value of the shift.
\end{abstract}


The Chern--Simons (CS) parameter shift was studied
by many authors from a perturbative point of view.
The shift is caused by the quantum correction,
but the outcome varies with the way of regularization.
Some authors conclude
that the CS parameter $\theta$ shifts by quadratic casimir $c_v$,%
~\cite{shift-results,Alvarez-Lbastida-Ramallo90}
and some others say that there is no shift.%
~\cite{non-shift-results}
Recently,
several authors~\cite{both-shift} showed that
the above ambiguity comes from 
the large-momentum behavior of the regularized action,
and we can get both the results
by a suitable choice of the regulators.

In the CS gauge theory
the treatment of the anti-symmetric symbol
$\epsilon^{\mu\rho\nu}$ is nontrivial.
As it is difficult to define $\epsilon^{\mu\rho\nu}$
in a critical dimension,
like $\gamma_5$ in four dimensional gauge theory,
the dimensional regularization
which is most convenient and popular for gauge theory
is problematic here.
%
%
In such a case,
Pauli--Villars (PV) regularization method is useful
because it does not touch the dimension.
In particular, when it is employed with 
higher covariant derivative (HCD) regulators
it is well known that the shift result is derived
from the one-loop corrections.%
\cite{Alvarez-Lbastida-Ramallo90}
On the other hand,
the no-shift result does not given by any PV method.
If there is an ambiguity depending on the regulator,
we ought to get the no-shift result
following the same procedure
which gives a shift result~\cite{Alvarez-Lbastida-Ramallo90}
by a suitable choice of regulators.

The hybrid regularization method consists of following steps.%
~\cite{Faddeev-Slavnov91}
First we introduce the HCD regulators
in the theory.
They improve the behavior of propagators at large momentum,
rendering the theory less divergent
at the cost of the emergence of new vertices,
and the theory is reduced to superrenormalizable,
i.e. there are just a finite number of divergent loops.
When we introduce
\begin{equation}
S_{\rm HCD}={\Lambda^{-(n+1)}}\int d^3x FD^{n}F
\end{equation}
as HCD action, 
the superficial degree of divergence is calculated as
$\omega = 3 - (n+1)(L-1) - E_A$,
where $F$ is field strength of gauge,
$n$ an integer parameter, 
$E_A$ the number of external gauge field
and $L$ the number of loop momentum.
All graphs except one-, two- and three-point functions
at one-loop level are convergent for $n \ge 1$.
We deal with the remainder by a PV type of regularization.

We have a choice of PV regulators to regulate the one-loop graphs.
The PV regulators are constructed from the original action
adding a mass term of parity-even.~\cite{Faddeev-Slavnov91}
In this case,
because the CS term has the parity-odd nature
the PV fields are the mixed states of even and odd in parity.%
~\cite{Alvarez-Lbastida-Ramallo90}
In actual calculation,
the cross terms from parity-even and -odd parts remain
at the large momentum limit
and they contribute to the shift.
Namely, the parity-breaking of PV fields causes the shift.
Then if we can make a parity-invariant regulators
which are uniform in parity,
all of them are parity-even or -odd,
any cross terms do not arise
and we expect that we can get the no-shift results.
We show that we can construct such regulators
introducing two types of PV fields.

In the following discussion
we consider $R^3$ with Euclidean metric $g_{\mu \nu}$
of signature $(+,+,+)$,
and the ${\rm SU}(N)$ gauge field 
is denoted as $A_\mu = A_\mu^a T^a$,
where $T^a$ is the anti-hermitian generator of the Lie algebra.
We take the structure constant $f^{abc}$
totally antisymmetric in their indices, with
$[T^a,T^b]=f^{abc} T^c$
and
${\rm Tr} (T^a T^b) = \frac{1}{2}\delta^{ab}$.
Following the chiral invariant regularization in chiral theory%
~\cite{chiral}
by introducing the infinitely many PV fields,
we write the regularized generating functional as follows:
\begin{align}
Z[J]&=
\int \mathcal{ D}A_\mu\mathcal{ D}b\mathcal{ D}c\mathcal{ D}\overline c
\exp
\left[-S_{\rm CS}-S_{\rm GF}-S_{\rm HCD}
\right]
\nonumber \\
&\times
\prod_{j=1}^\infty 
\left[
 \det{}^{-\frac{\alpha_j}{ 2}} \mathcal{ A}_j^+
\right]
\left[
 \det{}^{-\frac{\alpha_{-j}}{ 2}}\mathcal{ A}_{-j}^-
\right]
\prod_{\substack{ i=-\infty \\ i\ne 0}}^\infty
\Big[ 
        \det{}^{\gamma_i} \mathcal{ C}_i
\Big],
\label{eq:generating functional}
\end{align}
where $S_{\rm CS}$, $S_{\rm GF}$ and $S_{\rm HCD}$ are the
CS, gauge fixing and higher covariant derivative
actions.
The determinants
$\Big[\det{}^{-\frac{\alpha_j}{ 2}} \mathcal{ A}_j^\pm\Big]$ and
$\Big[\det{}^{\gamma_i} \mathcal{ C}_i\Big]$
are the PV field for gauge and for ghost, respectively.
The details will be given below.

We take the actions
\begin{align}
 S_{\rm CS} 
  &= -{\rm i} 
  \epsilon^{\mu \rho \nu}
  \int_x
   \frac{1 }{ 2} A_\mu \partial_\rho A_\nu
   + \frac{1 }{ 3!} g A_\mu A_\rho A_\nu,
 \label{eq:CS action org}
\\ 
 S_{\rm GF}
 &= 
 \int_x
  b\frac{\xi_0 }{ 2f^2(\partial / \Lambda)} b
  - b(\partial^\mu A_\mu)
  + \overline c(\partial^\mu D_\mu c),
 \label{eq:GF action}
\\
 S_{\rm HCD}
 &=
 -\frac{\rm i}{2\Lambda^2} \epsilon^{\mu \rho \nu}
 \int_x
  F_{\mu\lambda}D_\rho F_{\nu}^\lambda,
 \label{eq:HCD action}
\end{align}
where $c$, $\overline c$ and $b$ are
the ghost, anti-ghost and auxiliary fields, respectively,
$\xi_0$ the gauge fixing parameter,
and we take the CS parameter $\theta=4\pi/g^2$ positive.
The integration $\int_x\equiv\int d^3x$ is over the whole $R^3$.
The function $f(\partial / \Lambda)$
must behave higher than $(\partial / \Lambda)^n$
to ensure the convergence of higher loops in arbitrary gauge,
and we take a polynomial of $\partial / \Lambda$
such that $f(\partial / \Lambda)\rightarrow 1$
as $\Lambda \rightarrow \infty$.

The PV determinant for gauge field is
\begin{equation}
\left[\det{}^{-\frac{\alpha_j}{ 2}}\mathcal{ A}^\pm_j\right]
=\int\mathcal{ D}A_j^\pm{}_\mu
\exp\left[-S_{M_j}^\pm-S_{b_j}^\pm\right],
\end{equation}
where
\begin{gather}
S_{M_j}^\pm = 
\frac{1}{ 2} 
\int_x\int_y
A_j^\pm{}_\mu(x)
\left[
 \frac{\delta^2 \left(\pm S_{\rm CS}+S_{\rm HCD} \right)}
 {\delta A_\mu(x) \delta A_\nu(y)}
 - M_j g^{\mu\nu} \delta^3(x-y)
\right]
A_j^\pm{}_\nu(y),
\label{eq:usual PV}
\\
S_{b_j}^\pm = 
\int_x
\left[
 b_j \frac{\xi_j}{ 2 f^2} b_j
 -b_j (D^\mu A_j^\pm{}_\mu)
\right].
\label{eq:auxiliary field for PV}
\end{gather}
$b_j$ and $\xi_j$ are the `$b$-field' and
the `gauge fixing parameter' 
for PV field $A_j{}_\mu^\pm$ respectively.
We chose $\xi_j = M_j$ in the below
for the simplification of the calculation.
\footnote{%
$b_j$ and $\xi_j$ are introduced 
for the complete regularization of Yang-Mills theory.%
~\cite{Asorey-Falceto95}
When we introduce them
the Feynman rules of PV fields become similar to ones of gauge field,
and we can easily calculate the summation with index $j$.
}
To ensure the parity invariance
we introduce two types of PV regulator,
$A_j^+{}_\mu$ and $A_{-j}^-{}_\mu$,
which exchange under the parity transformation
$(x_1,x_2,x_3) \rightarrow (-x_1,x_2,x_3)$,
\begin{align}
A_j^+{}_\mu \rightarrow  A_{-j}^-{}_\mu \qquad
A_{-j}^-{}_\mu \rightarrow A_j^+{}_\mu.
\end{align}
Under the parity transformation
$\epsilon^{\mu\nu\rho}$ changes its sign
and the product of the PV determinants
$\left[\det{}^{-\frac{\alpha_j}{ 2}}\mathcal{ A}_j^+\right]
 \left[\det{}^{-\frac{\alpha_{-j}}{ 2}}\mathcal{ A}_{-j}^-\right]$
is invariant on the condition $M_j=M_{-j}$.
A pair of two PV fields
$A_j^+{}_\mu$ and $A_{-j}^-{}_\mu$
conserves parity
i.e. we have to introduce a `pair' of two PV fields
with the same index $j$
to construct a parity invariant regulator.

Consider when we regulate the theory
by the parity-invariant PV pairs.
A pair is made from two fields,
introducing one pair corresponds to
subtracting double the divergence.
Then to remedy the over subtraction
we introduce another parity-invariant pair of the opposite statistics,
which means adding double the divergence.
To remedy the over addition
we have to introduce the third pair.
We repeat such steps alternately until the divergence is canceled.
Namely,
we cannot regulate the theory by a finite number of PV fields,
but we need an infinite number
to regulate by the parity-invariant PV fields.
This is the reason why we introduced an infinite number of PV fields
in (\ref{eq:generating functional}).

Taking account of these facts,
the PV conditions are written as follows:
\begin{equation}
M_j=M|j|,\qquad
\alpha_j=(-1)^j.
\label{eq:PV conditions}
\end{equation}
As we see later,
the condition for $M_j$ not only guarantees the parity invariance
but leads to a well convergent function
with the help of the condition for $\alpha_j$,
when we calculate the quantum correction explicitly.
The condition for $\alpha_j$ means 
that we introduce fermionic PV field $(-1)$ and
bosonic one $(+1)$ alternately.
On this condition with an infinite number,
the summation of PV fields becomes
$-1+1-1+\cdots$ in an actual calculation
and it also recovers the usual PV condition
$1+\sum\alpha_j = 0$
where we may take $\alpha_j$ and the number of PV fields arbitrarily.
The infinite number is the cost
we are asked to pay for choosing the
alternating PV fields.

For ghost field
there is no parity-odd term in the action
and we use the following determinants for PV fields
with the same PV conditions $m_i = m|i|,\gamma_i = (-1)^i$:
\begin{equation}
\det\mathcal{ C}_i=\int
\mathcal{ D}\overline c^{(i)} \mathcal{ D}c^{(i)}
\exp
 \int_x
 \left\{
  \overline c^{(i)}D_\mu D^\mu c^{(i)}
  -m_i^2 \overline c^{(i)} c^{(i)}
 \right\},
\label{eq:detCi}
\end{equation}
which is parity-invariant by itself.

The regularized action (\ref{eq:generating functional})
is invariant under the BRST transformation
\begin{equation}
\begin{array}{cccc}
\delta_{\rm B} A_\mu = D_\mu c,&
\delta_{\rm B} b = 0,&
\delta_{\rm B} c = \{c,c\},&
\delta_{\rm B} \overline c = b,\\
\delta_{\rm B} A_j{}_\mu = [A_j{}_\mu,c ],&
\delta_{\rm B} b_j = [b_j, c],&
\delta_{\rm B} \overline c_i = \{\overline c_i,c\},&
\delta_{\rm B} c_i = \{c_i,c \}.
\end{array}
\end{equation}
Here we chose a homogeneous BRST transformation for the PV fields
to preserve the gauge invariance.~\cite{Alvarez-Lbastida-Ramallo90}

\begin{figure}[ht]
\begin{center}
\includegraphics{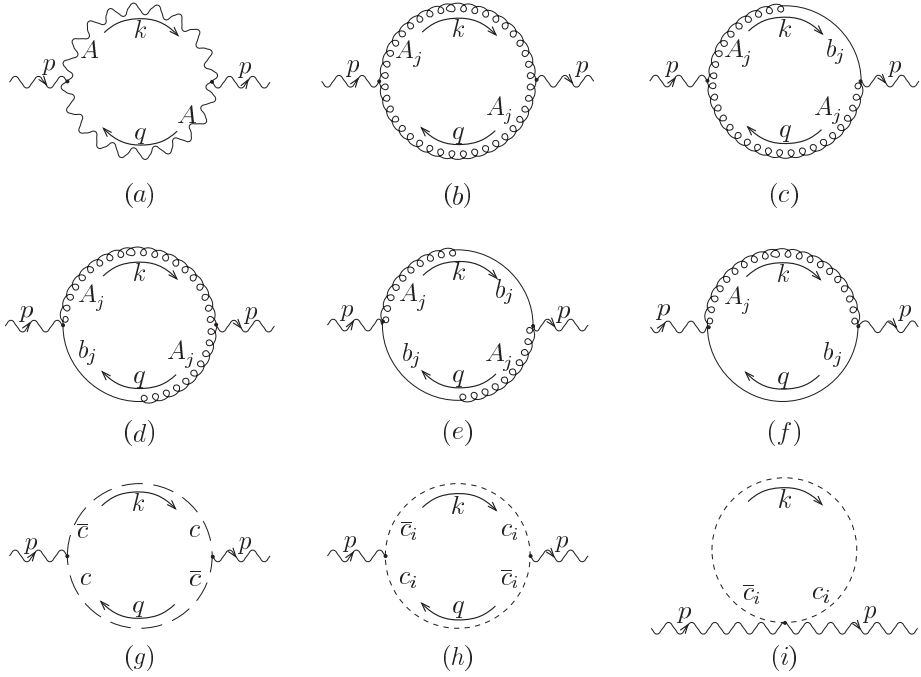}
\caption
{The loops contribute to the vacuum polarization tensor.
 The wavy line means the gauge field $A$,
 the curly line the PV field $A_j$,
 and the straight line the auxiliary field $b_j$.
 The rough- and fine-dotted line mean
 ghost $c$ and PV for ghost $c_i$, respectively.
 We use the same assignment of internal momenta, where $q=k-p$.
 $(a)$, $(b)$, $(c)$ and $(d)$ have both parity-even and
 parity-odd contributions.
 }
\label{fig:graph}
\end{center}
\end{figure}

Now we accomplish the task
to construct parity and gauge invariant regularization,
we are ready to calculate the quantum correction.
For simplicity
we consider the two-point functions in the following.
All the loops we have to calculate are listed
in Fig.~\ref{fig:graph}.

As our main aim is to calculate so-called `$\theta$--shift',
we concentrate to discuss on parity-odd terms
of the vacuum polarization tensor.
Each of the loops $(a)$, $(b)$, $(c)$ and $(d)$ in Fig.~\ref{fig:graph}
contains parity-odd term,
and the parity odd contributions come from these four loops.
First we consider the loops $(a)$ and $(b)$.
They have the same form,
and when we write the contribution of loop $(a)$
to the vacuum polarization tensor 
as ${}^{(a)}\Pi^{ab}_{\mu\nu}(p)$ and so on,
we get
\begin{align}
{}^{(a)}\Pi^{ab}_{\mu\nu}(p)\Big|_{\rm odd}&=
\int\frac{d^3k}{ (2\pi)^3}
\frac{g^2c_v\delta^{ab} }{ k^2q^2}
\left[
 \frac{\xi_0}{ q^2}
  \epsilon^{\nu\mu\rho}q_\rho qk
 +\frac{\xi_0}{ k^2}
  \epsilon^{\mu\nu\rho}k_\rho kq
\right],
\label{eq:(a)}
\\ \nonumber \\
{}^{(b)}\Pi_j{}^{ab}_{\mu\nu}(p)\Big|_{\rm odd} &=
\int\frac{d^3k}{ (2\pi)^3}
\frac{{\rm sign}(j) g^2c_v\delta^{ab} }
     { (k^2+M_j^2)(q^2+M_j^2)}
\bigg [
 M_j\epsilon^{\mu\nu\rho}p_\rho +
\nonumber \\
 &\quad
 \frac{\xi_j-M_j }{ (\xi_jM_j+q^2)}
  \epsilon^{\nu\mu\rho}q_\rho qk
 +\frac{\xi_j-M_j }{ (\xi_jM_j+k^2)}
  \epsilon^{\mu\nu\rho}k_\rho kq
\bigg ],
\label{eq:(b)}
\end{align}
where $q = k-p$ and $p$ is the external momentum.
Since (\ref{eq:(b)}) is proportional to the sign of $j$,
two contributions from $A^+_j$ and $A^-_{-j}$ have the same value
apart from the sign
and cancel each other 
\begin{align}
\sum_{j=1}^\infty (-1)^j
{}^{(b)}\Pi_j{}^{ab}_{\mu\nu}(p)
\Big|_{\rm odd}
+\sum_{j=-\infty}^{-1} (-1)^j
{}^{(b)}\Pi_j{}^{ab}_{\mu\nu}(p)
\Big|_{\rm odd}= 0.
\label{eq:vacb}
\end{align}
On the other hand,
from an explicit calculation of (\ref{eq:(a)})
we obtain ${}^{(a)}\Pi^{ab}_{\mu\nu}(p)\big|_{\rm odd}=0$
independently of $\xi_0$.

Next we consider the loops $(c)$ and $(d)$.
The mechanism of the cancellation is just the same as $(b)$.
These loops give the same value except sign,
i.e.
$
{}^{(c)}\Pi_j{}^{ab}_{\mu\nu}(p)
\big|_{\rm odd}
+
{}^{(c)}\Pi_{-j}{}^{ab}_{\mu\nu}(p)
\big|_{\rm odd}= 0$.
There is no contribution from these loops.
As a result, 
the parity odd terms are not induced 
through the vacuum polarization tensor.
We conclude that $\theta$ does not shift
with parity-invariant regulators.


To complete our regularization scheme
we also consider the parity-even contributions
from all the loops in Fig.~\ref{fig:graph}.
They contain divergent terms
which must be regularized.
An explicit calculation of ghost loops,
$(g)$, $(h)$ and $(i)$, shows
\begin{align}
{}^{(g)+(h)+(i)}\Pi^{ab}_{\mu\nu}(p)\Big|^{\rm div}
&=
g^2c_v\delta^{ab}
\int\frac{d^3k}{(2\pi)^3}
\frac{3k_\mu k_\nu-2k^2g_{\mu\nu}}{k^2q^2}
\nonumber \\
&+
g^2c_v\delta^{ab}
\sum_{i=-\infty}^{\infty}\gamma_i\int\frac{d^3k}{(2\pi)^3}
\frac{2(k^2+m_i^2)g_{\mu\nu}-4k_\mu k_\nu}
     {(k^2+m_i^2)(q^2+m_i^2)}.
\label{eq:divergent from ghost}
\end{align}
The first term originates
from the difference of Feynman rule expressions
between ghost and PV for ghost
and does not depend on mass parameter $m$.
In the second term
we regard the original ghost field as the zeroth PV field,
$c^a_0 \equiv c^a$
and write the summation with index $i$
from $-\infty$ to $\infty$.
The summation is reduced to the following two functions:
\begin{align}
\sum_{i=-\infty}^\infty (-1)^i
\int \frac{d^3k}{(2\pi)^3}
\frac{k_\mu k_\nu}{(k^2+m_i^2)(q^2+m_i^2)}
&=
\frac{g_{\mu\nu}}{16}
\left(\frac{m^4}{12p^3}-\frac{mC}{6\pi} + O(m^0)\right),
\\
\sum_{i=-\infty}^\infty (-1)^i
\int \frac{d^3k}{(2\pi)^3}
\frac{m_i^2}{(k^2+m_i^2)(q^2+m_i^2)}
&=
\frac{-1}{16}
\left(\frac{m^4}{12p^3}-\frac{mC}{6\pi} + O(m^0)\right), 
\end{align}
where $C$ is an arbitrary constant.
Using these results,
we find that in (\ref{eq:divergent from ghost})
the second term vanishes
and only the first term remains divergent.

This mechanism arises also in the calculation of the loops
from $(a)$ to $(f)$;
the divergent parts containing the mass parameter vanish
after the infinite sum with index $j$ from $-\infty$ to $\infty$.
By choosing $\xi_j = M_j$,
we calculate the divergent part as follows:
\begin{align}
{}^{(a)\sim (f)}\Pi^{ab}_{\mu\nu}(p)\Big|^{\rm div}_{\rm even}
&=
-g^2c_v\delta^{ab}
\int\frac{d^3k}{(2\pi)^3}
\frac{3k_\mu k_\nu-2k^2g_{\mu\nu}}{k^2q^2}.
\end{align}
The divergence is the same as the one in
(\ref{eq:divergent from ghost})
except for the sign
and the total divergence from the parity-even terms vanishes. 
What is shown hereto is that the theory is regularized completely,
that is, the regularization works properly.


In this note,
we consider the non-abelian CS gauge theory 
with the hybrid regularization
which consists of the HCD action
and the parity invariant PV fields,
and calculate the vacuum polarization tensor.
%
On the PV condition (\ref{eq:PV conditions}),
the parity-odd contribution from the
vacuum polarization tensor of $A^+_j$ and that from $A^-_j$
cancel out
because of the parity invariance. 
From these facts we can conclude that the CS shift does not occur.

We can derive the shift result 
introducing the non-parity-invariant PV regulators.%
~\cite{Alvarez-Lbastida-Ramallo90}
In this choice,
the parity-odd contributions from the PV loops,
like the first term of (\ref{eq:(b)}),
induce to the shift
since the cancellation mechanism of them does not work.
These two results show the ambiguity of the CS shift by the regulators.
Using the HCD term like $FDF$, instead of $DFDF$, 
we can show that the CS shift does occur.
It is considered that
each result belongs to the different universality class
which is classified by the parity of the regulators
at a large momentum.~\cite{both-shift}

%

Finally we comment on the choice of the HCD terms.
Alvarez-Gaum\`e {\it et al}.\cite{Alvarez-Lbastida-Ramallo90}
chose the HCD action as $DFDF$
and calculated the one-loop graphs
in the limit of $\Lambda \rightarrow \infty$.
This calculation is not acceptable
because the finite contribution is not the same as one
that we calculate in the finite $\Lambda$.
%
%
The inclusion of
$DFDF$ term modifies the Feynman rules drastically, 
the gauge propagator of the CS gauge theory
\begin{equation}
\frac{1}{p^2}\epsilon_{\mu\rho\nu}p^\rho
-\frac{\xi_0}{p^4}p_\mu p_\nu
\end{equation}
is modified to
\begin{equation}
\frac{\Lambda^6}{p^2(\Lambda^6+p^6)}
\epsilon_{\mu\rho\nu}p^\rho
-\frac{\xi_0}{p^4f^2(p)}p_\mu p_\nu
+\frac{\Lambda^3}{(\Lambda^6+p^6)}(p^2g_{\mu\nu}-p_\mu p_\nu).
\label{eq:modified propagator}
\end{equation}
%
The third term of (\ref{eq:modified propagator})
gives a finite contribution to the odd part of the vacuum polarization tensor
$\Pi^{ab}_{\mu\nu}(p)\big|_{\rm odd}$ with finite $\Lambda$.
%
Since this term makes the ghost self-energy nonzero,
the theory needs the renormalization procedure
which affects the shift.
In fact,
we can show the non-integer shift with $DFDF$ term directly.
%
This shift contradicts with the quantization condition%
~\cite{Deser-Jackiw-Templeton}
caused by the large-gauge transformation.
But if we add Yang-Mills term $FF$ to $DFDF$
we get the integer shift~\cite{Nittoh-Ebihara97}
in agreement with the result by usual methods.%
~\cite{shift-results}
This fact says that 
the naive treatment of $\Lambda$ may lead an incorrect result.

%


\end{document}